\documentclass[journal]{IEEEtran}
\usepackage[cmex10]{amsmath}
\usepackage{amssymb}
\usepackage{amsthm}
\usepackage{epsfig}
\usepackage{color}
\usepackage{algorithm}
\usepackage{algorithmic}
\usepackage{multirow}
\usepackage{array}
\newtheorem{theorem}{Theorem}

\usepackage{epsfig}
\usepackage{color}

\newcommand{\squeezeup}{\vspace{-2.5mm}}

\begin{document}
\title{Structural Vulnerability of Power Grids to Disasters:\\ Bounds, Adversarial Attacks and Reinforcement}
\author{Deepjyoti Deka,\IEEEmembership{~Student Member,~IEEE} and Sriram Vishwanath,\IEEEmembership{~Senior Member,~IEEE}
\thanks{Deepjyoti Deka and Sriram Vishwanath are with the Department of Electrical and Computer Engineering, The University of Texas at Austin, Austin, TX 78712 USA. This work is supported by the Defense Threat Reduction Agency (DTRA) through grant \# HDTRA1-09-1-0048-P00003. (e-mail:deepjyotideka@utexas.edu; sriram@ece.utexas.edu)}}

\maketitle
\begin{abstract}
 Natural Disasters like hurricanes, floods or earthquakes can damage power grid devices and create cascading blackouts and islands. The nature of failure propagation and extent of damage is dependent on the structural features of the grid, which is different from that of random networks. This paper analyzes the structural vulnerability of real power grids to impending disasters and presents intuitive graphical metrics to quantify the extent of damage. Two improved graph eigen-value based bounds on the grid vulnerability are developed and demonstrated through simulations of failure propagation on IEEE test cases and real networks. Finally this paper studies adversarial attacks aimed at weakening the grid's structural resilience and presents two approximate schemes to determine the critical transmission lines that may be attacked to minimize grid resilience. The framework can be also be used to design protection schemes to secure the grid against such adversarial attacks. Simulations on power networks are used to compare the performance of the attack schemes in reducing grid resilience.
\end{abstract}

\section{Introduction}
The topological structure of the power grid is an important feature that affects the delivery of electricity \cite{dekaISGTeigen}. From an economic perspective, the capacity of transmission lines and the graphical properties (whether tree-like or loopy) affect the locational marginal electricity prices as well as the convergence of Optimal power flow algorithms \cite{opfduality}. Insufficient line capacities and network structure can lead to highly fluctuation prices and even negative prices \cite{negativeprices}. From an reliability perspective, the grid structure influences the extent of damage following an natural or man-made disaster. In particular, it affects the propagation of failures after an initial breakdown of equipment and thus in turn affects the formation of islands and loss of service. Over the years, natural disasters like earthquakes, floods, hurricanes have caused extensive power outages due to damage of grid equipment and loss of network connectivity \cite{katrina,hurricaneindices}. Such disasters also affect co-located interdependent transportation and communication infrastructure as well. Thus there is a greater need to quantify the effect of the grid structure on failure propagation in the grid following a natural disaster and to incorporate the insights gained into transmission planning techniques to improve grid resilience.

There is existing work that studies the impact of structure on grid reliability. Reference \cite{dobson} describes a widely accepted and realistic DC model for the propagation of equipment (nodes and links) failures in cascading outages in the power grid. Here, the propagation process begins with an initial failure of a network node or link which leads to a redistributing of the power flows for optimal dispatch. This can lead to some transmission lines running above their prescribed capacities and subsequently tripping due to overheating. Subsequently in \cite{dobson1}, the authors incorporate recovery mechanism for the tripped transmission lines into the failure model and show that the size of blackout has a power-law distribution as seen in reality. Reference \cite{beinstock} analyzes the problem of finding the optimal $k$ lines in the grid that can be used for interdiction in the grid to create failures. Power Flow based analysis has been done to analyze geographically correlated failures in \cite{beinstock1}. Similarly, an interdiction based analysis on grid resilience considering short term impacts is discussed in \cite{yezhou}. However, these models include solving a Optimal Power Flow (OPF) or similar optimization problem to study the propagation of failures. Such an approach is harder to analyze. In particular, it seldom leads to a closed form expression of a metric or parameter that can quantify the resilience of the power grid to failures. In a separate line of work, efforts have been made to study probabilistic failure propagation in power grids and related networks using techniques from percolation theory and random graph theory. In this approach, initial failures are supposed to propagate probabilistically from source to neighboring nodes and edges in the grid graph before terminating. The final state will often include greater number of failures than the initial state and efforts are made to study the effect of the grid structure in influencing the spread. References \cite{albert2000, betweenness} study the effect of removing nodes from power grid graphs based on their centrality and degree measures and its effect on network connectivity. The authors of \cite{failure} analyze the propagation of structural failures in complex random networks based on similar neighborhood propagation rules. This approach has been extended to study interdependent networks failures as well \cite{interdependent} where nodes of two different networks depend on one another for survivability. Similarly, references \cite{stochasticgeo1, stochasticgeo2} have analyzed node and edge percolation based techniques to understand failure propagation in random graphs generated by stochastic geometry. An interaction graph based model is presented in \cite{interaction} where power flow based cascading information is used to generate an interaction graph for the network to study the inter-nodal dependencies on cascades. A good review of works pertaining to grid resilience to natural disasters can be found in \cite{review}.

It is worth mentioning that parallel analytical techniques are also used in studying social, biological and cyber-networks for information dissemination and spread of viruses \cite{miller2007,epidemiccorr}. However, the accuracy of percolation based techniques and further of the use of random graphs in modeling power grid cascading failures is debatable \cite{hinesdebate}. Existing work \cite{wang2010, deka, deka1} has demonstrated that the structure of real power grids as well as their finite sizes create significant deviations in observed graph parameters from those predicted in random graphs. This is because popular random network models like Erdos-Renyi, Barabasi-Albert, small world and configuration models \cite{complexgrid} do not accurately capture the specific nature of the spatio-temporal evolution of power grids. Further, sharp breakdown thresholds emerge in analysis of topological failure models on random graphs that are seldom observed in simulations of failures on real grid graphs and IEEE test cases \cite{test}. Such thresholds arise due to the absence of local loops and locally tree-like nature of random graph models that are not encountered in real grids. Hence, it is fair to suggest that analysis of random graphs to study the failure propagation (both probabilistic and power flow based) will not extend directly to real grids.

In this work, we focus on power grid failures induced by large natural disasters like hurricanes and earthquakes that create disconnected islands in the grid and loss of connectivity. We study the size of the largest connected component in the post-disaster grid and provide justification for using this as a valid metric for grid vulnerability in modern power grids and micro-grids that have non-trivial fraction of renewable and other distributed generation resources. Note that prior literature includes the use of the largest connected graph component in simulation based studies of grid failures \cite{mitigation}. This is distinct from failure propagation models where a failed node is considered to affect neighboring nodes with a degree or physical characteristic based probability. We extend probabilistic analysis previously used for random graphs on known real grid graphs and popular IEEE test cases to determine computable graphical parameters (Eg. eigenvalues of the grid adjacency matrix) that can be used to quantize the resilience of grids to such natural disasters. More importantly, we present a modified graph construction based on the true grid graph and use it to develop improved bounds on the extent of damage created in the network by the disaster \cite{dekaISGTeigen}. The efficacy of the graphical parameter based bounds and soft thresholds are demonstrated by simulations of failures on publicly available grid data-sets. We then use the graphical metrics on grid resilience to identify critical transmission lines (graph edges) that maximally affect the grid resilience. In particular, we study attack on grid resilience by an adversary that aims to damage a set of transmission lines to maximize the expected damage to network connectivity following a natural disaster. As this problem is NP-hard in general, we present two approximate algorithms to determine the optimal edges in the adversary's target set. The first algorithm is based on perturbation based analysis of eigen-values of the grid adjacency matrix while the second algorithm is based on greedy trace minimization of a higher power of the adjacency matrix. The performance of our algorithms for attack design in reducing grid resilience is demonstrated through simulations and also compared with other techniques in literature, notably attacks on nodes with high betweenness \cite{betweenness} or random attacks. From the system operator or grid controller's perspective, these algorithms can be used to determine the critical lines that need to be protected to build resilience and prevent further degradation of grid resilience before any impending natural disaster. To summarize, our work presents a analytical framework to quantify the resilience of real power grid graphs to natural disasters and develops two algorithms to determine the critical transmission lines that need to be protected to improve grid resilience and prevent adversarial deterioration.

The rest of the paper is organized as follows. In the next section, we develop our intuitive graph theoretic quantification of network resilience that is reasonable in the presence of local generation. Next, we analyze network failures and resilience in actual grid graphs without employing any assumptions from random graph theory in Section \ref{sec:bounds}. In Section \ref{sec:improvedbounds}, we present our novel modified graph construction and use it to develop improved bounds on size of the network damage along with simulation results on IEEE test cases and real power grids. We study adversarial attacks on transmission lines aimed at weakening grid resilience to natural disasters in Section \ref{sec:adversary} and present our approximate greedy methods to determine the critical transmission lines. Simulation results on our designed algorithms and comparison with existing work is presented in Section \ref{sec:simulation}. Finally, we discuss the insights gained and prospective future work in Section \ref{sec:conclusion}.

\section{Failure Model in Power Grids}
\label{sec:system}
We begin by describing the power grid model and its features.

\begin{figure}
\centering
\includegraphics[width=0.45\textwidth, height = .31\textwidth]{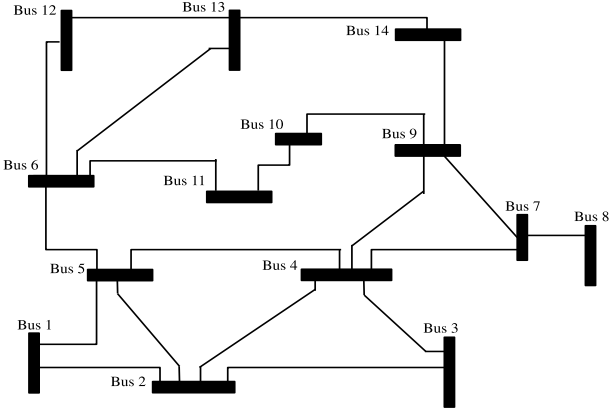}
\caption{IEEE 14-bus test system \cite{test}}
\label{fig:14bus}
\end{figure}

\textbf{Network Model:} We consider a modern power grid (or micro grid) in this paper that has distributed generation resources available on interior buses. Such generation may be provided by renewables (solar, wind etc.) or by conventional resources. We denote the grid by a graph ${\cal G} = (V, E)$, where sets $V$ and $E$ represent the nodes/buses and the undirected edges/lines respectively. Let the total number of buses in the system be $N$. We assume that under normal operating conditions, the lines have sufficient transmission capacity to transfer power from one part of the network to another. We denote the adjacency matrix of the graph $\cal G$ by $A_{\cal G}$ that is assumed to be known and not generated by a probabilistic model. Each edge $(ij)$ in $E$ is represented by a value of $1$ for $A(i,j)$ and $A(j,i)$ in the binary adjacency matrix. As an example, the IEEE $14$ bus test system \cite{test} is given in Figure~\ref{fig:14bus}.

\begin{figure}
\centering
\includegraphics[width=0.38\textwidth, height = .20\textwidth]{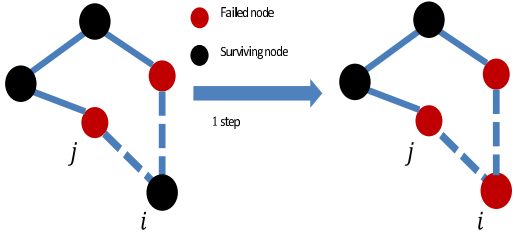}
\squeezeup
\squeezeup
\caption{Initial Node Failure ($j$) and Secondary Node Failure ($i$)}
\label{failure}
\end{figure}

\textbf{Failure Model:} As described in the Introduction, we consider a natural disaster that causes equipment failures in the grid. We assume that the natural disaster produces a probabilistic failure on all nodes in the system, with independent \emph{initial probability of failure} denoted by $p_0$. In this work, we only consider cases where the initial probability of failure on all nodes is the same. However, the entire analysis can be extended to cases where different nodes suffer distinct probabilities of initial failure. The initial failure rate $p_0$ depends on the nature of the natural disaster (Eg. earthquake scale, wind speed of hurricane etc.) as well as on the geographical placement of the grid (Eg. topography of the land will affect the failure rate). Such probabilities, in practice, are computed by agencies like the National Hurricane Center and used to predict the scale of damage and help in planning for evacuation strategies \cite{vaidy}. As we are concerned with connectivity in the network, we consider \emph{secondary failures} in surviving nodes that get separated from the rest of the network due to initial failures in all of their neighboring nodes. This is shown in Figure \ref{failure}. Next we describe our measure of network damage following the natural disaster.

\textbf{Network Damage}: As mentioned earlier, we assume that transmission line capacities are sufficient to satisfy all load in the system provided enough generating resources are online and connected. Let $N_s$ denote the size of the largest connected component of surviving nodes after the disaster. We consider $N- N_s$ to be the \emph{measure of network damage} caused by the natural disaster. It is worth mentioning that outside of the largest component, smaller groups of nodes can be functional as well if enough generation is available to satisfy the total load in the node group. We select $N- N_s$ as the measure of network damage as it has a key characteristic as described next.

Let $\Delta^P_i$ denote the net power capacity (generating capacity minus load) of node $i$ in the network, where nodal generating capacity and load are both random variables. Consider the case where each $\Delta^P_i$ is an independent Gaussian random variable with mean $\mu \geq 0$ and variance $\sigma^2$. If node $i$ is disconnected from the rest of the network, the probability $q^i_0$ that its local load is served is then given by:
\begin{align}
&q^i_0 = \mathbb{P} (\Delta^P_i \geq 0) = \int_{0}^{\infty}\frac{1}{\sqrt{2\pi\sigma^2}}e^{-(x - \mu)^2/\sigma^2}dx\nonumber\\
\Rightarrow &q^i_0 = .5 + \int_{0}^{\mu}\frac{1}{\sqrt{2\pi\sigma^2}}e^{-(x - \mu)^2/\sigma^2}dx \quad\text{as~}\mu \geq 0\label{1node}
\end{align}
On the other-hand, if node $i$ is connected to a group of $N_k$ nodes, the probability $q^i_{N_k}$ that node $i$ satisfies its load is given by:
\begin{align}
q^i_{N_k} &= \mathbb{P} (\sum^{N_k}_{i=1}\Delta^P_i \geq 0) = \mathbb{P} (\sum^{N_k}_{i=1}\Delta^P_i/N_k \geq 0)\nonumber\\
&= .5 + \int_{0}^{\mu}\frac{1}{\sqrt{2\pi\sigma^2/N_k}}e^{-\frac{(x - \mu)^2}{\sigma^2/N_k}}dx\label{nsnode}
\end{align}
where Eq.~(\ref{nsnode}) follows from the fact that $\sum^{N_k}_{i=1}\Delta^P_i/N_k$ is a Gaussian$(\mu,\sigma^2/N_k)$ random variable. Note that its variance decreases with increase in $N_k$, the size of the connected set that node $i$ belongs to. For $1\leq N_{k}\leq N_{s}$, using properties of the exponential function it follows that
\begin{align}
&\int_{0}^{\mu}\frac{1}{\sqrt{2\pi\sigma^2/N_{k}}}e^{-\frac{(x - \mu)^2}{\sigma^2/N_{k}}}dx \leq \int_{0}^{\mu}\frac{1}{\sqrt{2\pi\sigma^2/N_{s}}}e^{-\frac{(x - \mu)^2}{\sigma^2/N_{s}}}dx\label{temp1}
\end{align}
Using Eqs.~(\ref{1node}), (\ref{nsnode}) and (\ref{temp1}), we have $q^i_{1} \leq q^i_{N_{k}}\leq q^i_{N_{s}}$.

Thus, \emph{the probability of a nodal load being served increases with an increase in the size of the connected component that the node belongs to.} Thus the largest component shows the highest group of nodes whose cumulative loads are satisfied with highest probability. This justifies our usage of $N_s$ (size of the largest connected component of surviving nodes) to quantify the functional network and correspondingly of $N-N_s$ to measure the scale of network damage. In the next section, considering the largest component as the surviving network, we analyze the effects of network structure on the extent of failures and determine a preliminary upper bound on the probability of initial failure $p_0$ beyond which the network fragments.

\section{Failure Analysis and Preliminary Bound}
\label{sec:bounds}
As shown in Fig.~\ref{failure}, we consider initial failures and secondary failures in the grid and analyze their creation in discrete steps. Let $\lambda^V_t$ denote the vector of survival probabilities of all $N$ nodes in set $V$ at step $t$. For node $i$, we have $\lambda^V_0(i) = 1-p_0$ where $p_0$ is the initial probability of failure. According to the failure model, node $i$ survives at step $t$ if it did not fail at step $t=0$ and did not get disconnected between steps $1$ and $t-1$. In other words, at least one of its neighbors did not fail by step $t-1$. We express this mathematically as
\begin{align}
&\lambda^V_t(i) = (1-p_0)\mathbb{P}[\bigcup_{j:(ij)\in {E}}\{\text{node $j$ survives at $t-1$}\}]\nonumber\\
\Rightarrow~&\lambda^V_t(i) \leq (1-p)\sum_{j:(ij)\in {E}}\lambda^V_{t-1}(j)\label{next}\\
\Rightarrow~&\lambda^V_t \leq (1-p)A_{\cal G}\lambda^V_{t-1} \label{bound1}
\end{align}
Here Eq.~(\ref{next}) follows from the Union Bound for probabilities. Note that for a general graph $\cal G$, this gives an inequality as against an equality that is obtained for a random graph model \cite{interdependent} where failure propagation from each distinct neighbor is independent. This is a crucial distinction as real world power grid graphs are not locally tree-like and have correlated failure pathways unlike random graphs. Let $\beta_A$ be the largest eigenvalue of the adjacency matrix $A_{\cal G}$. Using relation (\ref{bound1}), we have
\begin{align}
(1-p_0)\beta_A < 1, \text{~then~}\lambda^V_{\infty} \rightarrow \textbf{0}\label{first}
\end{align}
Thus, $p_0 > 1-1/\beta_A$ provides a \textbf{upper bound} on the threshold on initial probability of random failures ($p_0$) beyond which the grid fragments. In contrast, random graph analysis leads to an exact threshold and not an upper bound. It is worth noting that the current formulation does not specify the extent of damage in the region $p_0 < 1-1/\beta_A$. In the next section, we present a novel modified graph construction that overcomes this and helps generate tighter bounds.

\section{Modified Graph for Improved Bounds}
\label{sec:improvedbounds}
Note that in our failure model, the survival of any node depends on the existence of edges connecting it to the largest connected component. Indeed we can analyze a node's survivability by considering the probability of it being connected through operational edges in the grid graph. To motivate this approach better, consider two connected neighboring nodes $i$ and $j$ as shown in Fig.~\ref{modifiedgraph}. Let $B^{E}_t(ij)$ be the event that node $i$ is connected to the surviving nodes in the largest component through edge $(ij)$ at step $t$. Let the probability of $B^{E}_t(ij)$ be denoted by $\lambda^{E}_t(ij)$. Note that this event can be defined for every neighboring node of node $i$ and $i$'s survivability requires at least one event to be true. Thus, the probability of node $i$ surviving at step $t$ is given by:
\begin{figure}
\centering
\includegraphics[width=0.28\textwidth, height = .18\textwidth]{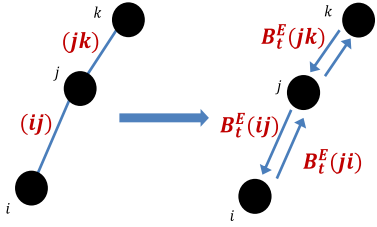}
\squeezeup
\squeezeup
\caption{Each undirected edge results in two survival probabilities, one in each direction}
\label{modifiedgraph}
\end{figure}
\begin{align}
\lambda^V_t(i) = (1-p_0)\mathbb{P}[\bigcup_{j:(ij)\in {E}}[B^{E}_t(ij)] \label{modified}
\end{align}
Here, the $(1-p_0)$ arises from the probability of node $i$ surviving an initial failure, while the remaining terms correspond to the survivability due to a connecting edge. In a similar way, we define event $B^{E}_t(ji)$ of probability $\lambda^{E}_t(ji)$ for node $j$ surviving through the edge with node $i$. This event is the reciprocal of $B^{E}_t(ij)$. Thus, every edge gives rise to two probabilities of survival, one along each direction as shown in Figure \ref{modifiedgraph}. If nodes $i$, $j$ and $k$ are connected as shown in Fig.~\ref{modifiedgraph}, the event $B^{E}_t(ij)$ (node $i$ surviving via edge $(ij)$) depends on $B^{E}_t(jk)$ (node $j$ surviving through edge $(jk)$). In terms of their probabilities, $\lambda^{E}_t(ij)$ and $\lambda^{E}_{t-1}(jk)$ are related. Extending to other nodes in the system, we write this relation mathematically $\forall i,j$ such that $(ij) \in {E}$ as
\begin{align}
&\lambda^{E}_t(ij) = (1-p_0)\mathbb{P}[\bigcup_{k:(jk)\in E, k\neq i}B^{E}_{t-1}(jk)]\nonumber\\
\Rightarrow~&\lambda^{E}_t(ij) \leq (1-p_0)\sum_{k:(jk)\in E,k\neq i}\lambda^{E}_{t-1}(jk)~~\text{(Union Bound)} \label{bound2}
\end{align}
In Eq.~(\ref{bound2}), the $(1-p_0)$ comes from the fact that if node $j$ fails initially, then node $i$ cannot survive through edge $(ij)$. Eq.~(\ref{bound2}) is the motivation behind our construction of a modified graph based on the power grid graph to better estimate the scope of network failures.

\begin{figure}
\centering
\includegraphics[width=0.36\textwidth, height = .20\textwidth]{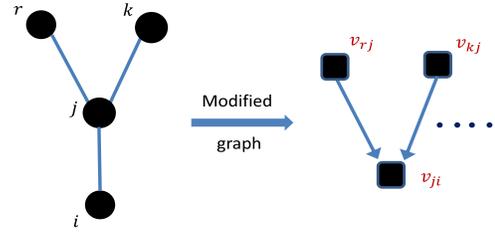}
\squeezeup
\squeezeup
\caption{Modified graph ${\cal G}^{E}$ formation from $\cal G$}
\label{modifiedgraph1}
\end{figure}

\textbf{Modified Graph Construction:} We create modified directed graph ${\cal G}^{E}$ from the original graph $\cal G$ as follows
\begin{itemize}
\item Each edge $(ij)$ in $E$ gives rise to two nodes $v_{ij}$ and $v_{ji}$ in ${\cal G}^{E}$.
\item Two nodes $v_{ab}$ and $v_{cd}$ in ${\cal G}^{E}$ are connected by an edge directed from $v_{ab}$ towards $v_{cd}$ if $b=c$ and $a\neq d$ (see Figure \ref{modifiedgraph1}).
\end{itemize}

Note that ${\cal G}^{E}$ is similar in structure to a line graph of grid graph $\cal G$. However, there is a slight difference in that ${\cal G}^{E}$ has lesser number of edges as a pair of nodes $v_{ab}$ and $v_{ba}$ are \textit{not} neighbors in ${\cal G}^{E}$, though they would be in a standard line graph construction. The number of nodes in ${\cal G}^{E}$ is equal to twice the number of edges in $\cal G$. The size of $A_{E}$, the adjacency matrix of ${\cal G}^{E}$, is $2|{E}|\times 2|{E}|$. We now write Eq.~(\ref{bound2}) in a vector form similar to Eq.~(\ref{bound1}) using the adjacency matrix of the modified graph ${\cal G}^{E}$ as follows:
\begin{align}
\lambda^{E}_t \leq (1-p_0)A_{E}\lambda^{E}_{t-1} \label{bound3}
\end{align}
Here each node in the modified graph is associated with one probability of failure that represents directional connectivity by the corresponding edge in the original graph $\cal G$. Extending the analysis in the previous section to Eq.~(\ref{bound3}), it follows that if $\beta_{A_{E}}$ (the largest eigenvalue of $A_{E}$) satisfies $p_0 > 1-1/\beta_{A_{E}}$, then the grid disintegrates as $\lambda^{E}_{\infty} \rightarrow \textbf{0}$ and $\lambda^V_{\infty} \rightarrow \textbf{0}$. Thus, $1-1/\beta_{A_{E}}$ provides a \textbf{second upper bound} on the threshold on $p_0$, beyond which the grid fragments. We compare the first upper bound $1-1/\beta_A$ in (\ref{first}) and the second lower bound $1-1/\beta_{A_{E}}$ of IEEE test-cases and real grids and observe that in all cases the second upper bound is smaller in magnitude and hence provides an improved tighter bound on the failure threshold. The comparisons are noted in simulation results later in this section.

We now use the modified graph ${\cal G}^{E}$ and Eq.~(\ref{modified}) to analyze the extent of damage in the grid. In particular, we are interested in bounding the size of number of failures over the entire range of initial probability of failure $p_0$, even below the bounds derived earlier. For this, we use the basic failure definition where a node fails eventually if either it fails initially or if it does not get connected to the largest component through any of its edges. In other words, the probability of \emph{not} surviving (given by $1-\lambda^{V}_t(ij)$ for node $i$ in graph $\cal G$) depends on the probability of initial failure and on the probability that \emph{none} of the neighbors of node $v_{ij}$ in ${\cal G}^{E}$ survives. Let $[B^{E}_t(ij)]^c$ denote the event of node $i$ \emph{not surviving} through edge $(ij)$. Mathematically, we have
\begin{align}
1-\lambda^V_t(i) &= p_0 +(1-p_0)\mathbb{P}[\bigcap_{j:(ij)\in {E}}[B^{E}_{t-1}(ij)]^c] \label{reason0}\\
&\leq p_0 +(1-p_0)\min_{j:(ij)\in {E}}(1 - \mathbb{P}[B^{E}_{t-1}(ij)]) \label{reason1}\\
&\leq p_0 +(1-p_0)\sum_{j:(ij)\in {E}}\frac{1 -\lambda^{E}_{t-1}(ij)}{d_i} \label{reason2}
\end{align}
where $d_i$ is the number of neighbors of node $i$ in $\cal G$. Eq.~(\ref{reason0}) follows from the failure definition where $p_0$ denotes the initial failure probability for original node $i$. Eq.~(\ref{reason1}) follows from $\mathbb{P}[A\cap B] \leq \min(\mathbb{P}[A],\mathbb{P}[B])$ while Eq.~(\ref{reason2}) follows from the fact that the minimum of a set of numbers is less than their average. Likewise, we express ($1-\lambda^{E}_t(ij)$) for failure probabilities in the modified graph as below.
\begin{align}
1-\lambda^{E}_t(ij) &= p_0 +(1-p_0)\mathbb{P}[\bigcap_{k:(jk)\in E, k\neq i}[B^{E}_{t-1}(jk)]^c]\\
&\leq p_0 +(1-p_0)\sum_{k:(jk)\in E, k\neq i}\frac{1-\lambda^{E}_{t-1}(jk)}{d_{ij}} \label{reason3}
\end{align}
where $d_{ij}$ is the number of neighbors of node $v_{ij}$ in modified graph ${\cal G}^{E}$. Writing it in vector form for $t\rightarrow \infty$, we get
\begin{align}
\textbf{1}-\lambda^{E}_{\infty} = p_0(\mathbb{I}_{2|E|}- (1-p_0)D_{E}^{-1}A_{E})^{-1}\label{part}
\end{align}
where $\mathbb{I}_{2|E|}$ is the identity matrix of dimension $2|E|$ (number of nodes in ${\cal G}^{E}$). $D_{E}$ is the diagonal matrix of node degrees ($d_{ij}$) in ${\cal G}^{E}$, while $A_{E}$ is its adjacency matrix. The upper bound on the expected number of node failures in the gird ($N_f$) is then given by:
\begin{align}
N_f &= \sum^N_{i=1}(1-\lambda^V_{\infty}(i))\leq p_0N + (1-p_0)\textbf{1}^TD_{\cal G}^{-1}A_{\cal G}(\textbf{1}-\lambda^{E}_{\infty})\nonumber\\
&\leq p_0N + (1-p_0)\sum^N_{i=1}\sum_{j:(ij)\in {E}}\frac{1 -\lambda^{E}_{\infty}(ij)}{d_i} \label{final}
\end{align}
Using Eqs.~(\ref{part}) and (\ref{final}), the \textbf{upper bound} on node failures can be computed. If there are nodes in $\cal G$ with unit degree (a common observation in power grids), the upper bound on number of failures will be non-trivial. To demonstrate the performance of our bounds, we present simulations of random failures on known power grid graphs.

\subsection{Comparison of Bounds through simulations}
\begin{figure}[h]
\centering
\includegraphics[width=0.45\textwidth]{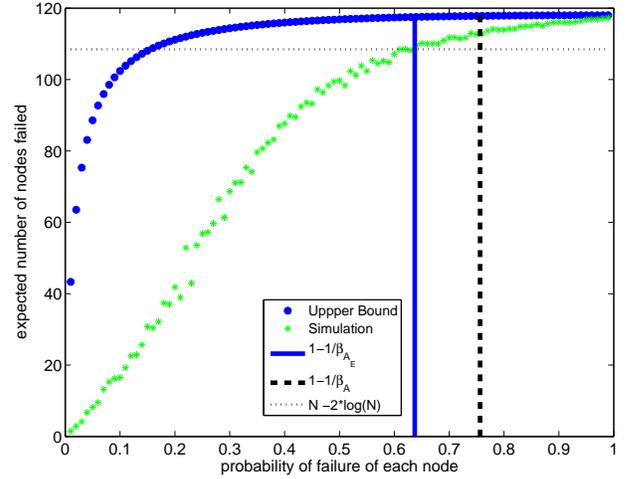}
\squeezeup
\caption{Upper bounds for number of failed nodes and bounds on $p_0$ in IEEE $118$ bus system}
\label{plot118}
\end{figure}

\begin{figure}[h]
\centering
\includegraphics[width=0.45\textwidth]{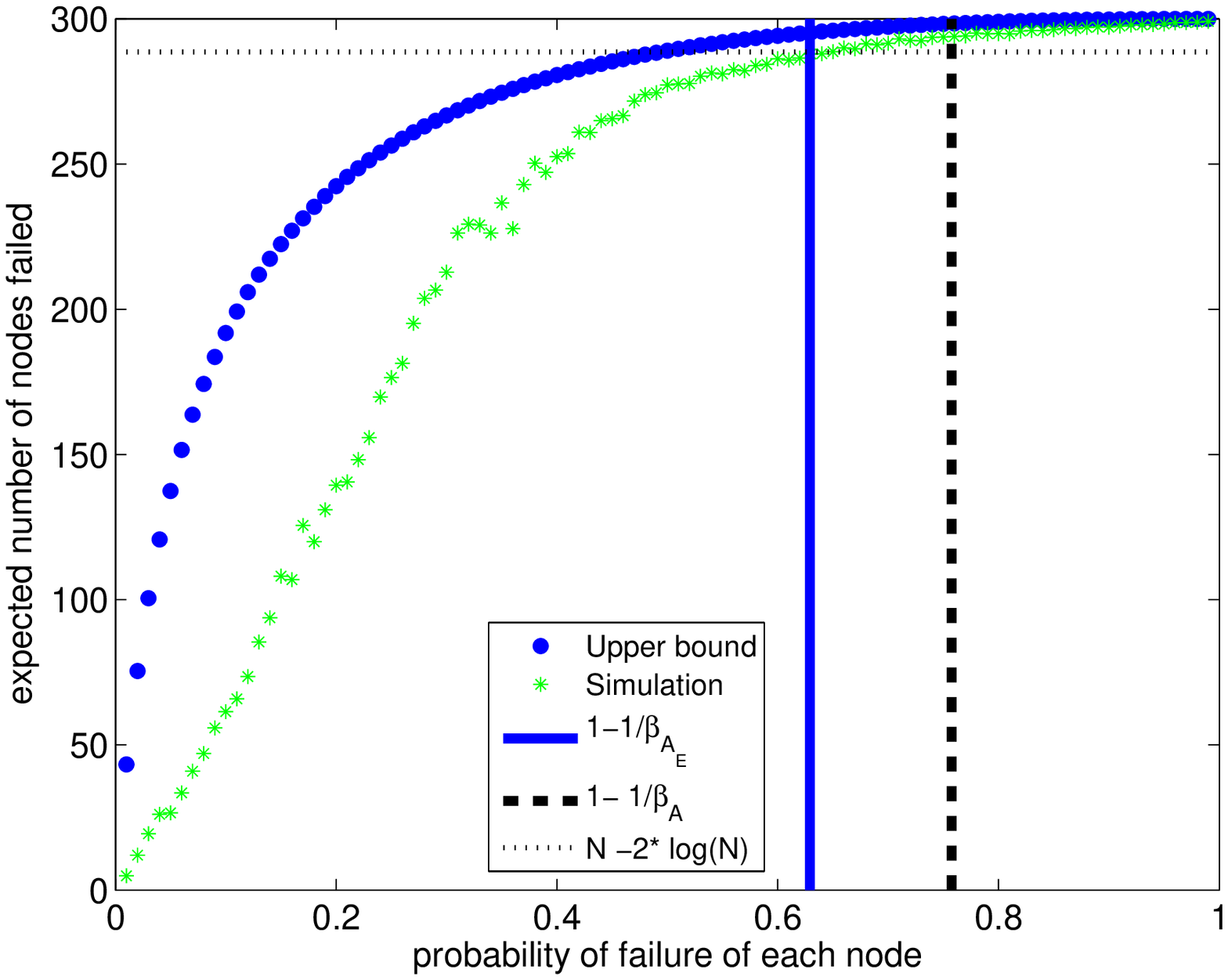}
\squeezeup
\caption{Upper bounds for number of failed nodes and bounds on $p_0$ in IEEE $300$ bus system}
\label{plot300}
\end{figure}

\begin{figure}[h]
\centering
\includegraphics[width=0.45\textwidth]{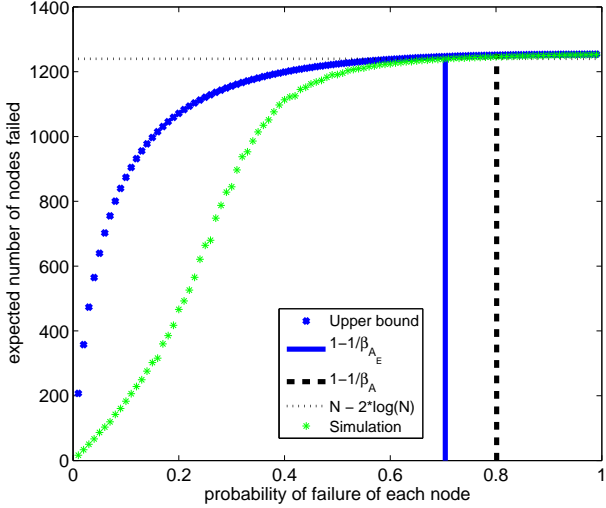}
\squeezeup
\caption{Upper bounds for number of failed nodes and bounds on $p_0$ in Western US power grid}
\label{plotwestern}
\end{figure}

\begin{figure}[h]
\centering
\includegraphics[width=0.45\textwidth,height = .36\textwidth]{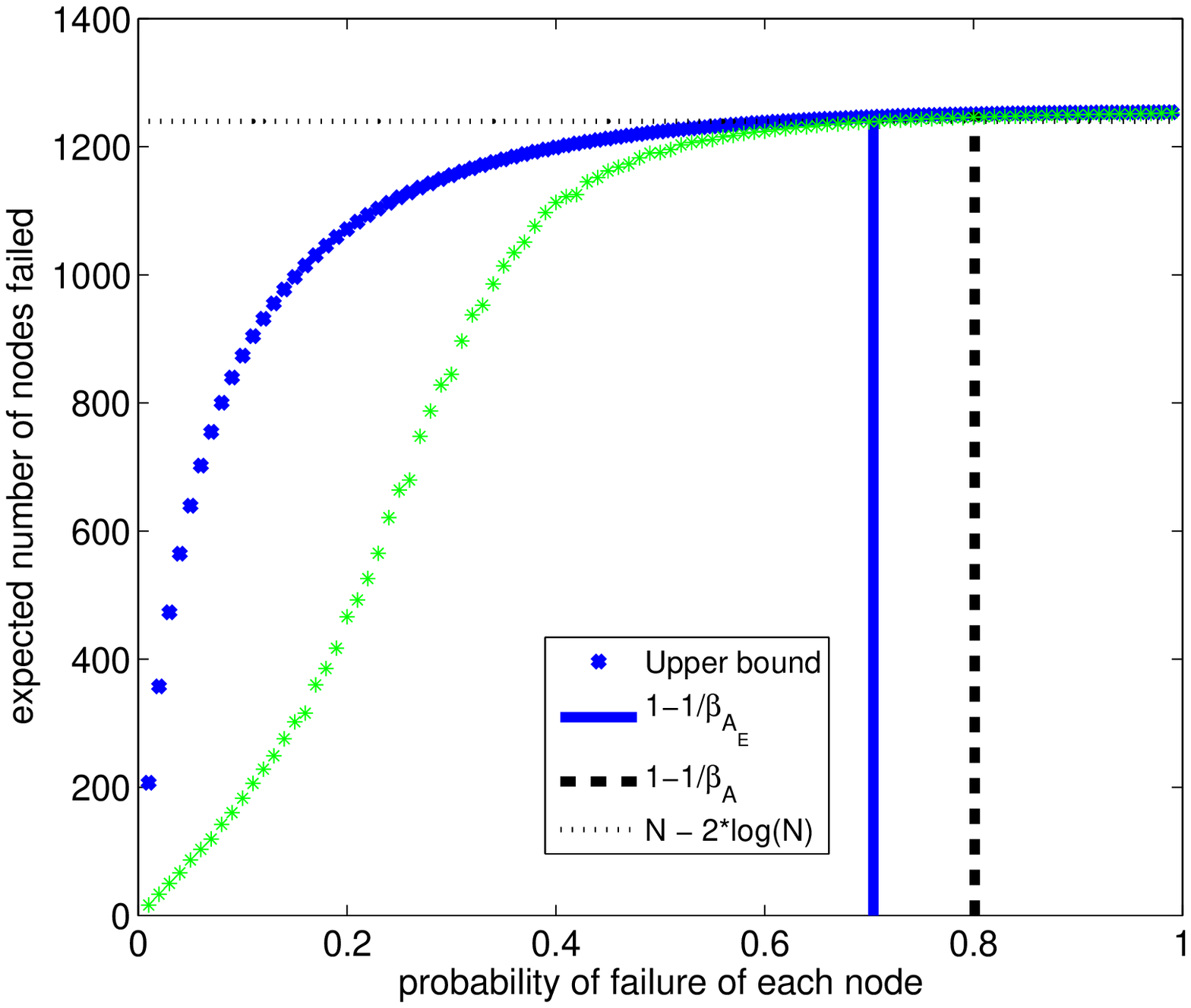}
\squeezeup
\caption{Upper bounds for number of failed nodes and bounds on $p_0$ in UCTE power grid}
\label{ploteurope}
\end{figure}

The performances of the upper bound on network failure over all values of $p_0$ and the two upper bounds on critical value of $p_0$ beyond which the network disintegrates are shown through simulations on the IEEE $118$ and $300$ bus test systems \cite{test} in Figs.~\ref{plot118} and \ref{plot300} respectively. Subsequently we also consider publicly available power grid topologies pertaining to the Western US grid and the grid under the Union for Coordination of Transmission of Electricity (UCTE) in Europe. The Western US grid has $4941$ nodes and $6594$ edges \cite{watts1998} while the power grid of the UCTE has $1254$ buses and $1811$ lines \cite{europe2005}. Failure propagation simulation and determined bounds for these networks are shown in Figs.~\ref{plotwestern} and \ref{ploteurope}.

In all the power grid cases, we consider the grid to have fragmented if the size of the largest connected component of surviving nodes is less than $2\log N$, where $N$ is the initial size of the network. Note that the second upper bound ($1-1/\beta_{A_E}$) on the threshold on $p_0$ (initial probability of failure) derived from the modified graph is lower and hence tighter than the first upper bound ($1-1/\beta_{A}$) derived using the original graph. Further, it needs to be pointed out that the upper bound on the number of failures given by Eqs.~(\ref{part}) and (\ref{final}) is tighter for higher values of $p_0$ compared to smaller values. The original graph does not provide a non-trivial upper bound on failures over the entire range of $p_0$, hence the modified graph has two distinct advantages in failure analysis.

In the next section, we discuss how our measures of network reliability based on eigenvalues can be used to determine critical transmission lines that may be attacked by an adversary interested in weakening the grid resilience to natural disasters.

\section{Critical Lines for Adversarial Attack on Grid Resilience}
\label{sec:adversary}
We consider an adversary that aims to maximally weaken the grid structure to make it more vulnerable to failures during natural disasters. The adversary does so by attacking and removing a fixed number ($k_{max}$) of transmission lines in the grid. As mentioned in prior sections, relations involving the eigenvalues of adjacency matrix of the grid graph or the modified graph provide upper bounds on the probability of failure beyond which grid connectivity diminishes greatly. In the remainder of this section, we focus our attention on the first upper bound $(1- 1/\beta_A)$ in Eq.~(\ref{first}), where $\beta_A$ is the largest eigenvalue of the adjacency matrix $A_{\cal G}$ of the grid graph $\cal G$. We thus formulate the adversary's objective as damaging $k_{max}$ edges in the graph to minimize $\beta_A$ to reduce the first upper bound. Techniques based on the second upper bound based on the modified graph in Eq.~(\ref{bound3}) will be the focus of our future work in this area.

Let normalized eigenvector $u_1$ correspond to the largest eigenvalue $\beta_A$ of adjacency matrix $A_{\cal G}$. By definition $u_1$ satisfies
\begin{align}
\max_{\|x\|_2=1} x^TA_{\cal G}x = u_1^TA_{\cal G}u_1 = \beta_A \label{eigen1}
\end{align}
The Perron-Frobenius theorem \cite{stewart} states that eigenvector $u_1$ is a positive vector. Using this fact with Eq.~(\ref{eigen1}), it is clear that removing edges from graph $\cal G$ (or deleting $1$s from $A_{\cal G}$) always leads to a reduction in the magnitude of its largest eigenvalue. Hence, \textbf{the adversary's attack consists of determining the critical edges that will enable the maximal reduction in the largest eigenvalue of $A_{\cal G}$.} This is a known NP-hard problem. Here, we present two approximate techniques to determine the critical lines that will be included in the adversary's target set.

\textbf{Eigen-Perturbation based Attack Design:} In this attack scheme, we use perturbation analysis \cite{stewart} to approximate the change in eigenvalues of adjacency matrix $A_{\cal G}$ following removal of edges and subsequently to determine the optimal transmission lines to attack. Let the new adjacency matrix of the grid after removal of edges be given by $A_{\cal G}- \Delta A_{\cal G}$. The change in adjacency matrix $\Delta A_{\cal G}$ has the following structure:
\begin{align}
\Delta A_{\cal G}(i,j)= \begin{cases}1&~~\text{if edge $(ij) \in E$ is removed},\\
0 &~~\text{otherwise,} \end{cases} \label{delta}
\end{align}
Let the largest eigenvalue of new adjacency matrix be $\beta^{\Delta}_{A}$. From Eq.~(\ref{eigen1}), we have
\begin{align}
&\beta^{\Delta}_{A} = \max_{\|x\|_2=1} x^T(A_{\cal G}- \Delta A_{\cal G})x\nonumber\\
\Rightarrow &\beta^{\Delta}_{A} \geq {u_1}^T(A_{\cal G}- \Delta A_{\cal G})u_1 = \beta_A -{u_1}^T\Delta A_{\cal G}u_1\nonumber\\
\Rightarrow &\Delta \beta_A = \beta_A - \beta^{\Delta}_A\approx
 \sum_{\text{removed~}(ij)} 2u_1(i)u_1(j) (\text{~~using} (\ref{delta})) \label{trick1}
\end{align}
where $\Delta \beta_A$ denotes the change in the maximum eigen-value following the removal of lines.

\textbf{The optimal transmission line whose removal approximately minimizes the maximum eigenvalue of the adjacency matrix is thus given by maximizing the expression in Eq.~(\ref{trick1}).} To determine the optimal $k_{max}$ lines, the adversary iteratively computes $u_1$, the eigenvector corresponding to the largest eigenvalue, removes line by maximizing Eq.~(\ref{delta}) and recomputes the adjacency matrix and its principal eigenvector. As shown later, the iterative scheme provides a far greater reduction in grid resilience than selecting the $k_{max}$ lines to maximize Eq.~(\ref{delta}) all at once, though it leads to an increase in computational complexity.

\textbf{Complexity:} The computation of the eigenvector $u_1$ takes $O(N^3)$ steps via Singular Value Decomposition (SVD) of $A_{\cal G}$ \cite{matrixtheory}. Given that the maximization of Eq.~(\ref{trick1}) takes $|E|$ steps which is less than $O(N^3)$, computing the $k_{max}$ critical lines by iteratively computing the eigenvector for the largest eigenvalue has a complexity of $O(k_{max}N^3)$\cite{matrixtheory}. If all lines are selected based on a single computation of $u_1$, the complexity is $O(N^3)$ due to SVD. Next we describe another technique for attack design that depends on trace minimization.

\textbf{Trace Minimization based Attack Design:} The trace of a matrix refers to the sum of its diagonal elements and is equal to the sum of its eigenvalues \cite{stewart}. Consider an even $2r^{th}$ power of the trace of the adjacency matrix $A_{\cal G}$. As the eigenvalues of $A_{\cal G}^{2r}$ are the $2r^{th}$ powers of the eigenvalues ($\beta_1=\beta_A,\beta_2,...,\beta_{N}$) of $A_{\cal G}$, we have the following relation for the trace
\begin{align}
&trace(A_{\cal G}^{2r}) = \sum_{i = 1}^N \beta_i^{2r} = \beta_A^{2r}(1 + (\frac{\beta_2}{\beta_A})^{2r} +... + (\frac{\beta_N}{\beta_A})^{2r}) \label{trace1}\\
\Rightarrow~& trace(A_{\cal G}^{2r})/\beta_A^{2r}\approx 1 ~\text{as~} r \rightarrow \infty\label{trick3}
\end{align}
Here $\beta_A = \beta_i$ is the largest eigenvalue of the adjacency matrix. Note that as $|\frac{\beta_2}{\beta_A}| < 1$, if we take higher values of $r$, the ratio of eigenvalues becomes smaller in Eq.~(\ref{trace1}). Thus for extremely large values of $r$, the largest eigenvalue and trace of the $2r^{th}$ power of $A_{\cal G}$ are approximately equal as noted in Eq.~(\ref{trick3}). In this approach, thus we focus on reducing the trace of $A_{\cal G}^{2r}$ by removing lines instead of minimizing the largest eigenvalue $\beta_A$ or its higher power. Finding the optimal set of $k_{max}$ edges to minimize the trace of $A_{\cal G}^{2r}$ is computationally hard as well, however using trace minimization has certain advantages as we discuss now.

\begin{theorem}\label{submodopt}
The trace of $A_{\cal G}^{2r}$, where $A_{\cal G}$ is the adjacency matrix of grid graph $\cal G$ is a supermodular function of the constituent edges in the graph.
\end{theorem}
A real-valued function $f$ defined over set $S$ is supermodular \cite{submodular} if $f(A \bigcup C) \geq f(B \bigcup C)$ for $B \subset A$ and $A,B,C$ are subsets of $S$. In other words, the returns due to addition of $C$ are not diminishing.
\begin{proof}
Note that the $i^{th}$ diagonal element in $A_{\cal G}^{2r}$ is equal to the number of cycles of length $2r$ that begin and end at node $i$. This can be shown by direct checks or by mathematical induction. Here, cycle of length $2r$ refers to a graph path with $2r$ hops (repetition allowed) that begins and ends at the same node. Thus, the trace (sum of the diagonal elements of $A_{\cal G}^{2r}$) is given by the total number of cycles of length $2r$ that can be formed on all nodes in the grid graph. To show supermodularity of trace of $A_{\cal G}^{2r}$ as a function of graph edges, it is sufficient to show that the increase in the number of cycles of length $2r$ in graph $\cal G$ after adding a new edge $(ij)$ is less than the increase observed if edge $(ij)$ is added \emph{after inclusion of another edge} $(lm)$. This increase is indeed true as presence of an edge $(lm)$ prior to the addition of edge $(ij)$ will permit the existence of additional cycles that includes both edges $(lm)$ and $(ij)$, and cannot exist without $(lm)$. Hence trace of higher power of the adjacency matrix is a supermodular function of the graph edges.
\end{proof}

It is a known property \cite{submodular} that greedy minimization of a supermodular function is equivalent to greedy maximization of a submodular function and is provably at least $1-1/e$ ($\approx63\%$) close to the optimal solution. Thus, the adversary's attack policy in this scheme is \textbf{to greedily remove $k_{max}$ edges that minimizes the trace of $2r^{th}$ power of the adjacency matrix of the grid graph.}

\textbf{Complexity:} The $2r^{th}$ power of the symmetric adjacency matrix is computed efficiently using Singular Value Decomposition (SVD) as $A_{\cal G}^{2r} = U\beta_A^{2r}U^T$ where columns of $U$ are the eigenvectors and $\beta_A^{2r}$ is the diagonal matrix with $2r^{th}$ powers of the eigenvalues. Note that matrix multiplication and SVD are computed in $O(N^3)$ while computing $\beta_A^{2r}$ takes complexity $O(N\log r)$. Since we greedily minimize the trace, the selection of one edge takes $O(|E|(N^3 + N\log r))$. The overall complexity of computing $k_{max}$ optimal edges by this scheme is thus $O(k_{max}|E|(N^3 + N\log r))$. This expression implies that increasing $r$ to improve the accuracy of this approach will at most lead to a logarithmical increase in the complexity.

\textbf{Resilience:} From the grid controller's perspective, these two techniques can be used to determine the critical transmission lines for enhancing security and reinforcement to prevent adversarial manipulation aimed at disrupting grid resilience to natural disasters. In the next section, we look at the performance of these two approaches as an adversarial tool.

\section{Simulation Results of Adversarial Attacks}
\label{sec:simulation}
\begin{figure}[h]
\centering
\includegraphics[width=0.45\textwidth,height = .36\textwidth]{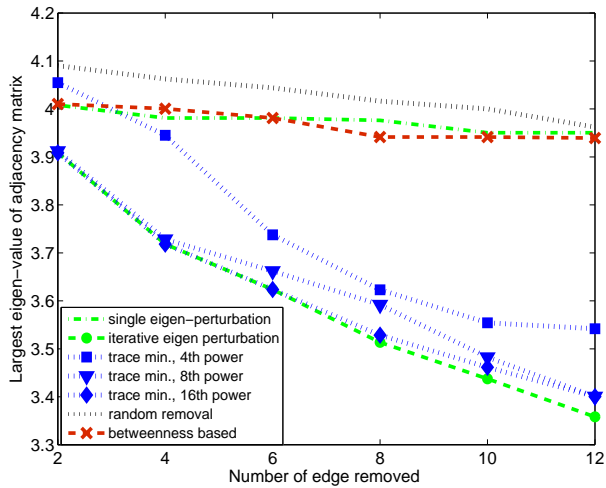}
\squeezeup
\caption{Comparison of adversarial schemes to reduce maximum eigenvalue of grid adjacency matrix in IEEE $118$ bus test system.}
\label{eigenchange118}
\end{figure}

\begin{figure}[h]
\centering
\includegraphics[width=0.45\textwidth,height = .36\textwidth]{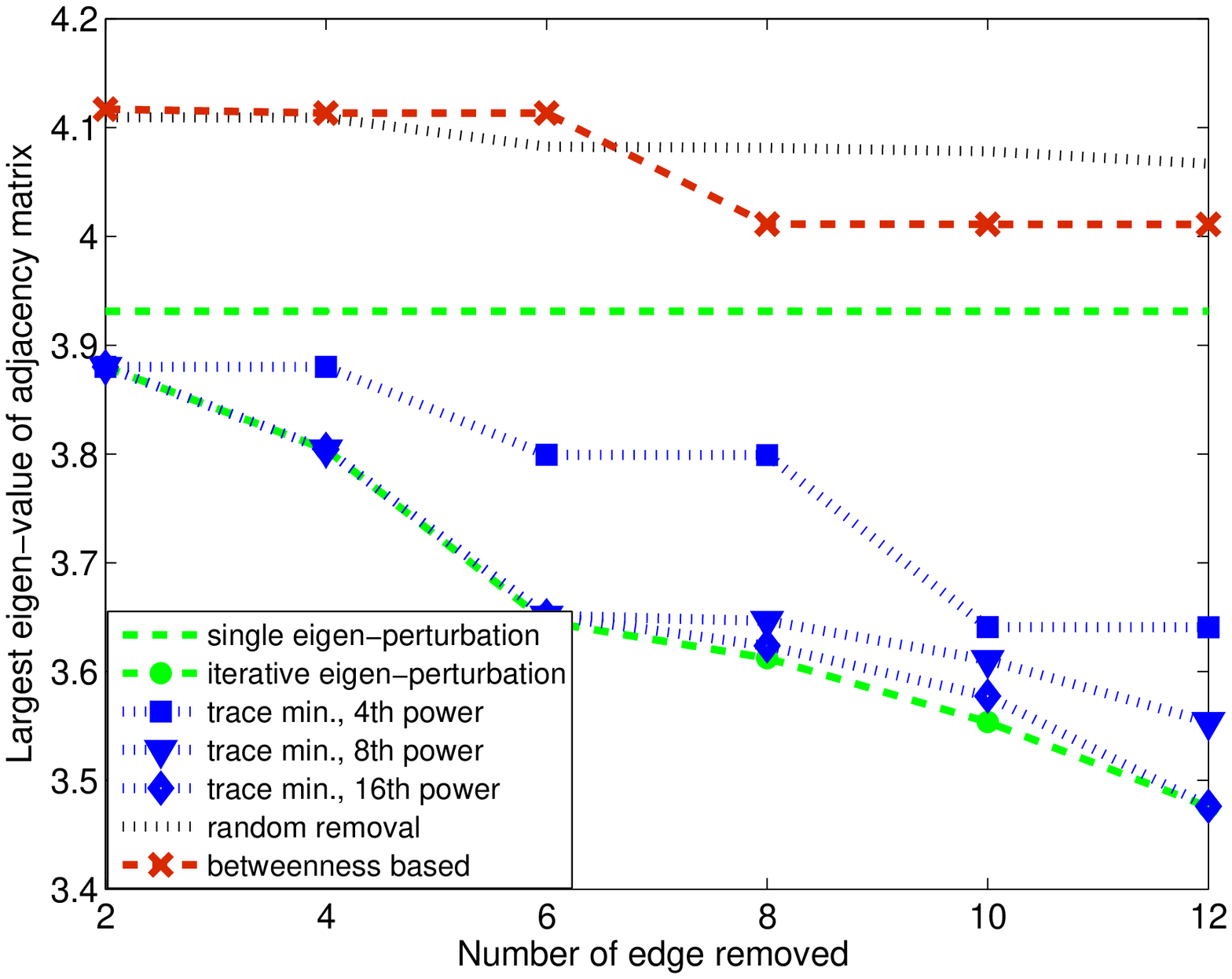}
\squeezeup
\caption{Comparison of adversarial schemes to reduce maximum eigenvalue of grid adjacency matrix in IEEE $300$ bus test system.}
\label{eigenchange300}
\end{figure}

\begin{figure}[h]
\centering
\includegraphics[width=0.45\textwidth,height = .36\textwidth]{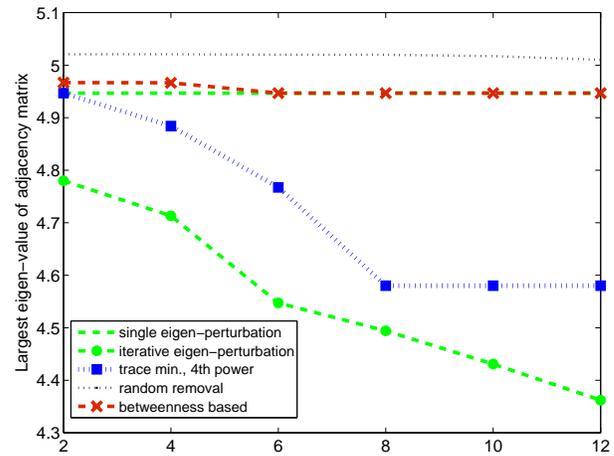}
\squeezeup
\caption{Comparison of adversarial schemes to reduce maximum eigenvalue of grid adjacency matrix in UCTE power grid.}
\label{eigenchangeeurope}
\end{figure}

We consider both approaches (eigen perturbation and trace minimization) for determining the optimal $k_{max}$ edges to minimize the largest eigenvalue of the adjacency matrix of the grid graph and thereby reduce the resilience of the grid to natural disasters. For comparison, we consider two alternate schemes, one where an adversary removes edges randomly, and another where an adversary removes edges in the decreasing order of their betweenness centralities \cite{betweenness}. We plot our results for the IEEE $118$ and $300$ bus test systems and the UCTE power grid network in Figs.~\ref{eigenchange118}, \ref{eigenchange300} and \ref{eigenchangeeurope} respectively. Note that both algorithms outperform random and betweenness based attacks to reduce the eigenvalues. It can also be noted that iterative eigen perturbation reduces the largest eigenvalue further than edge removal based on a single perturbation computation as mentioned in the previous section. Further, it can be observed from Figs.~\ref{eigenchange118} and \ref{eigenchange300} that increasing the value of $2r$, the power of the adjacency matrix, leads to an improvement in the trace minimization based scheme as it approximates the largest eigenvalue better as noted in Eq.~(\ref{trick3}).


\section{Conclusion}
\label{sec:conclusion}
We analyze topological vulnerability of power grids to probabilistic failures introduced by natural disasters in this paper. We present intuitive evidence that in modern grids and micro-grids where distributed generation resources are present, a reasonable metric of damage is given by the size of the largest connected component in the post-event grid graph. We analyze the evolving failure process that originates at nodes with initial failures. Based on the largest eigenvalue of the adjacency matrix of the grid, we present an upper bound on the critical probability of node failures beyond which the grid fragments. Further, we present the construction of a modified graph to analyze the probabilistic failures and use it to generate a tighter upper bound on the critical probability. This modified graph construction also enables us to derive new non-trivial upper bounds on the expected number of total failures for all values of the initial failure probability. We present the performance of our derived analytical bounds through simulations on two IEEE test cases and two real grid data sets. Finally, we discuss adversarial attacks on the power grid aimed at damaging transmission lines to minimize the grid's resilience to natural disasters. We develop two approximate algorithms to identify the critical lines that will enable such adversarial attacks. The first algorithm is based on perturbation analysis of the eigenvalues of the adjacency matrix and the second algorithm is based on greedy minimization of the trace of a higher power of the adjacency matrix. We analyze both algorithms and their complexity and demonstrate their performance against random and centrality based attacks studied in literature through simulations. Potential areas of future work include improving the bounds and developing a framework to incorporate topological analysis into power flow based studies on grid vulnerability to enhance its practical contribution.


\begin{thebibliography}{1}
\bibitem{dekaISGTeigen}
D. Deka, R. Baldick, and S. Vishwanath, ``Structural Vulnerability of Power Grids to Disasters: Bounds and Reinforcement Measures", {\em IEEE Conf. on Innovative Smart Grid Tech.}, 2015.
\bibitem{opfduality}
J. Lavaei and S. Low, ``Zero duality gap in optimal power flow problem", {\em IEEE Trans. Power Systems}, vol. 27, 2012.
\bibitem{negativeprices}
W. Chi-Keung, I. Horowitz, J. Moore, and A. Pacheco, ``The impact of wind generation on the electricity spot-market price level and variance: The Texas experience", {\em Energy Policy}, vol. 39, 2011.
\bibitem{katrina}
A. Kwasinski, W. W. Weaver, P. L. Chapman, and P.T. Krein, ``Telecommunications Power Plant Damage Assessment for Hurricane Katrina Site Survey and Follow-Up Result", {\em IEEE Systems Journal}, vol.3, 2009.
\bibitem{hurricaneindices}
L. Kantha, ``Time to Replace the Saffir-Simpson Hurricane Scale?", {\em Eos}, vol. 87, 2006.
\bibitem{dobson}
I. Dobson, B.A Carreras, V.E. Lynch,and D.E. Newman, ``An initial model fo complex dynamics in electric power system blackouts", {\em Proc. 34th Hawaii Int. Conf on System Sciences}, 2001.
\bibitem{dobson1}
I. Dobson, B.A Carreras, V.E. Lynch,and D.E. Newman, ``Complex systems analysis of series of blackouts: Cascading failure, critical points, and self-organization", {\em Chaos}, vol. 17, 2007.
\bibitem{beinstock}
D. Bienstock and A. Verma, ``The N − k problem in power grids: New models, formulations, and numerical experiments", {\em SIAM J. Optim.}, vol. 20, 2010.
\bibitem{beinstock1}
A. Bernstein, D. Hay, M. Uzunoglu, and G. Zussman, ``Power Grid Vulnerability to Geographically Correlated Failures: Analysis and Coontrol Implications", {\em arxiv preprint}, 2012. Avaible at: http://arxiv.org/abs/1206.1099.
\bibitem{albert2000}
R. Albert, H. Jeong, and A.-L. Barabasi, ``Error and attack tolerance of complex networks", {\em Nature}, vol. 406, 2000.
\bibitem{failure}
J. Wang and L. Rong, ``Cascade-based Attack Vulnerability on the US Power Grid", {\em Safety Science}, volume 47, 2009.
\bibitem{interdependent}
S. V. Buldyrev, R. Parshani, G. Paul, H. E. Stanley, and S. Havlin, ``Catastrophic cascade of failures in interdependent networks", {\em Nature}, vol. 464, 2010.
\bibitem{complexgrid}
G. A. Pagani and M. Aiello, ``The power grid as a complex network: a survey", {\em Physica A: Statistical Mechanics and its Applications}, 2013.
\bibitem{stochasticgeo1}
Z. Kong and E. M. Yeh, ``Resilience to degree-dependent and cascading node failures in random geometric networks", {\em IEEE Trans. Inf. Theory},
vol. 56, 2010.
\bibitem{stochasticgeo2}
H. Xiao and E. M. Yeh, ``Cascading link failure in the power grid: A percolation-based analysis" {\em Proc. IEEE Int. Work. on Smart Grid Communication}, 2011.
\bibitem{hinesdebate}
P. Hines, E. Cotilla-Sanchez, and S. Blumsack, ``Do topological models provide good information about electricity infrastructure vulenrablity?", {\em Chaos}, vol. 20, no. 3, p. 033122, Sept. 2010.
\bibitem{Durrett}
R. Durrett, ``Random graph dynamics", {\em Cambridge University Press}, 2006.
\bibitem{mitigation}
C. M. Schneider, A. A. Moreira, J. S. Andrade, S. Havlin, and H. J. Herrmann, ``Mitigation of malicious attacks on networks", {\em Proc. National Academy of Sciences}, 2011.
\bibitem{interaction}
J. Qi, K. Sun, and S. Mei, ``An interaction model for simulation and mitigation of cascading failures", {\em IEEE Trans. Power Systems}, 2015.
\bibitem{miller2007}
J. C. Miller and J. M. Hyman, ``Effective vaccination strategies for realistic social networks", {\em Physica A}, vol. 386, 2007.
\bibitem{epidemiccorr}
M. Boguna, R. Pastor-Satorras and A. Vespignan, ``Epidemic spreading in complex networks with degree correlations", {\em arXiv:cond-mat/0301149v1 [cond-mat.stat-mech]}, 2003.
\bibitem{wang2010}
Z. Wang, A. Scaglione, and R. Thomas, ``Generating statistically correct random topologies for testing smart grid communication and control networks", {\em IEEE Trans. Smart Grid}, vol. 1, 2010.
\bibitem{deka}
D. Deka and S. Vishwanath, ``Generative Growth Model for Power Grids", {\em Int. Conf. on Signal-Image Tech and Internet-Based Systems}, 2013.
\bibitem{deka1}
D. Deka and S. Vishwanath, ``Analytical Models for Power Networks: The case of the Western US and ERCOT grids", {\em arxiv preprint}, 2015. Available at: http://arxiv.org/abs/1204.0165.
\bibitem{yezhou}
Y. Wang and R. Baldick, ``Interdiction Analysis of Electric Grids combining cascading outage and medium-term impacts", {\em IEEE Trans. Power Systems}, 2014.
\bibitem{review}
Y. Wang, C. Chen, J. Wang, and R. Baldick, ``Research on Resilience of Power Systems Under Natural Disasters—A Review", {\em IEEE Trans. Power Systems}, 2015.
\bibitem{test}
University of washington, power systems test case archive. Available: http://www.ee.washington.edu/research/pstca.
\bibitem{vaidy}
V. Krishnamurthy and A. Kwasinski, ``Empirically validated availability model of information and communication technologies facilities under hurricane conditions", {\em IEEE INTELEC}, 2014.
\bibitem{europe2005}
Q. Zhou and J. W. Bialek, ``Approximate Model of European Interconnected System as a Benchmark System to Study Effects of Cross-Border Trades", {\em IEEE Transactions on Power Systems}, Vol. 20, No. 2, May 2005.
\bibitem{betweenness}
A. E. Motter and Y. -C. Lai, ``Cascade-based attacks on complex networks", {\em Phys. Rev. E}, vol. 66, 2002.
\bibitem{watts1998}
D. J. Watts and S. H. Strogatz, ``Collective dynamics of 'small-world' networks", {\em Nature}, vol. 393, 1998.
\bibitem{europe2005}
``Approximate Model of European Interconnected System", Available: http://www.see.ed.ac.uk/~jbialek/Europe load flow/.
\bibitem{matrixtheory}
G. H. Golub and C. F. V. Loan, {\em Matrix computations}, Vol. 3, JHU Press, 2012.
\bibitem{stewart}
G.W. Stewart and J. Sun, ``Matrix perturbation theory", 1990.
\bibitem{submodular}
G. Nemhauser, L. Wolsey, and M. Fisher, ``An analysis of the approximations for maximizing submodular set functions", {\em
Mathematical Programming}, vol. 14, 1978.
\end{thebibliography}
\end{document}